\documentclass[preprint,preprintnumbers,amsmath,amssymb]{revtex4}
\usepackage[colorlinks=true, pdfstartview=FitV, linkcolor=blue, citecolor=red, urlcolor=magenta]{hyperref}
\usepackage{amssymb}
\usepackage{latexsym}
\usepackage{epsfig}
\usepackage{color}
\allowdisplaybreaks

\begin{document}

\title{\bf Generalized Gravitational Baryogenesis of Well-Known $f(T,T_G)$ and $f(T,B)$ Models}
\author{Nadeem Azhar \footnote{nadeemazharsaeed@gmail.com}, Abdul Jawad
\footnote{jawadab181@yahoo.com;~~abduljawad@cuilahore.edu.pk},
 and Shamaila Rani
\footnote{drshamailarani@cuilahore.edu.pk}}
\address {Department of Mathematics, COMSATS University\\ Islamabad,
Lahore-Campus, Lahore-54000, Pakistan.}
\date{}
\begin{abstract}
The baryogenesis presents the theoretical mechanism that describes
the matter-antimatter asymmetry in the history of early universe. In
this work, we investigate the gravitational baryogenesis phenomena
in the frameworks of $f(T, T_G)$ (where $T$ and $T_G$ are the
torsion scalar and teleparallel equivalent to the Gauss-Bonnet term
respectively) and $f(T, B)$ (where $B$ denotes the boundary term
between torsion and Ricci scalar) gravities. For $f(T,T_G)$-gravity,
we consider two generic power law models while logarithmic and
general Taylor expansion models for $f(T,B)$-gravity. We consider
power law scale factor for each model and compute baryon to entropy
ratio by assuming that the universe filled by perfect fluid and dark
energy. We find generalized baryogenesis interaction which is
proportional to $\partial_\mu f(T+T_G)$ and $\partial_\mu f(T+B)$
for both theories of gravity. We compare our results against current
astrophysical data of baryon to entropy ratio, which indicates
excellent consistency with observational bounds (i.e.,
$\frac{\eta_B}{S} = 9.42 \times
10^{-11}$).\\\\


\end{abstract}

\maketitle

\section{Introduction}

The excess of matter over antimatter remains not only a biggest
puzzle in the history of early universe, but also an open problem in
modern cosmology. The observational data like measurements of cosmic
microwave background (CBM) \cite{1}, supported with big bang
nucleosynthesis \cite{2}, indicate more matter than antimatter in
the universe. Many authors presented a lot of theories to explore
this enigma, some of which are Affleck-Dine baryogenesis
\cite{3}-\cite{5}, electroweak baryogenesis \cite{6,7}, grand
unified theories (GUTs) \cite{8}, spontaneous baryogenesis
\cite{9}-\cite{11}, baryogenesis of thermal and black hole
evaporation \cite{12}, all these theories explain why there exists
matter antimatter asymmetry in our universe. Observational
constrains verify that the baryon number density to entropy ratio is
approximately $\frac{\eta_B}{S}\sim 9.42 \times 10^{-11}$ \cite{1,2}
where $\eta_B$ and $S$ denotes the number of baryon, and the entropy
of universe, respectively. Sakharov \cite{13} pointed out three
fundamental conditions which are needed to generate baryon
asymmetry. These conditions are

\begin{itemize}
  \item processes that
violate baryon number,
  \item violation of charge (C) and charge-parity (CP) symmetry,
  \item thermal inequilibrium.
\end{itemize}
Davoudiasl et al. \cite{14} proposed required matter-antimatter
asymmetry by the means of thermal equilibrium during transition
phase of universe while CP dynamically violated. The key ingredient
is a CP violating interaction which specified by coupling between
between the baryon matter current $J^\mu$ and the derivative of the
Ricci scalar curvature $R$, in the form
\begin{eqnarray}\label{111}
\frac{1}{M_*^2}\int{\sqrt{-g}d^4x(\partial_\mu R) J^\mu},
\end{eqnarray}
where $M_*$ characterizes the cutoff scale of the underlying
effective gravitational theory \cite{15}. In case of flat FRW
geometry $\frac{\eta_B}{S}\propto \dot{R}$, where overhead dot means
the derivative of $R$ with respect to time $t$. In case of radiation
dominated era whose equation of state $w=\frac{1}{3}$, the net
baryon asymmetry produced by Eq.(\ref{111}) tends to be zero.

Many authors extended baryogenesis phenomena in the framework of
modified theories gravity, which developed by modifying the Einstein
Hilbert action. In these theories of gravity, curvature-based
formulation of general relativity is the interesting and suitable
modification. However, teleparallel equivalent to general relativity
(TEGR) is another promising modification, in which curvature scalar
replaced by torsional formulation. Gravitational framework of this
theory, Lagrangian density support Weitzenb\"{o}ck connection
instead of the torsion-less Levi-Civita. Further generalization form
of this theory can be obtained by using general function $f(T)$
instead of torsion scalar $T$, namely $f(T)$-gravity. Hence,
similarly to the $f(T)$-gravity, one can construct $f(R)$ as a
extensions of TEGR by replacing curvature scalar $R$ instead of
Lagrangian density. $f(T)$ and $f(R)$ represent different
modification classes, therefore they do not coincide with each
other.

Beside this simple modification, one can construct more complicated
classes by introducing higher-torsion corrections just like
Gauss-Bonnet (GB) term $G$ \cite{15a}, Weyl combinations \cite{15b},
Love-lock combinations \cite{15c} etc. Based on this concept,
another modification of Einstein's theory presented known as
$f(T_G)$-gravity \cite{15d}. Hence by adding $f(T)$ term, another
generalization of $f(T_G)$-gravity presented known as $f(T, T_G)$
gravity. Recently, a latest modification of $f(T)$-gravity was
proposed by introducing a new Lagrangian $f(T,B)$, where $B$ is the
boundary term related to the divergence of the torsion tensor
($B=2\nabla_\mu (T^\mu)$). The $f(T,B)$-gravity \cite{15e} becomes
equivalent to $f(R)$ for the special choice $f(-T+B)$.

Nojiri and Odintsov \cite{nl1} reviewed various modified theories of
gravity and found that these theories have quite rich cosmological
structure. These theories demonstrated effective late-time era
(cosmological constant, quintessence or phantom) with a possible
transition from deceleration to acceleration and may pass the solar
system tests. Same authors \cite{nl2} discussed the general
properties and different representations of string-inspired and
Gauss-Bonnet theory, $f(R)$-gravity and its modified form, nonlocal
gravity, scalar-tensor theory, power-counting renormalizable
covariant gravity. Felice and Tsujikawa \cite{nl3} worked on dark
energy, inflation, cosmological perturbations, local gravity
constraints and spherically symmetric solutions in weak and strong
gravitational backgrounds by consider $f(R)$-gravity. Various well
known dark energy models for different fluids are explicitly
realized, and their properties are also explored \cite{nl4}. They
found these dark energy universes may mimic the $\Lambda$CDM model
currently, consistent with the recent observational data also paid
special attention to the equivalence of different dark energy
models. Nojiri et al. \cite{new} discussed some astrophysical
solutions and their several qualitative features in the framework of
modified theories of gravity. They emphasized on late-time
acceleration of universe, inflation, bouncing cosmology and formed a
virtual toolbox, which cover all necessary information about these
cosmological terms. However, Oikonomou \cite{new1} investigated how
the baryogenesis phenomena can potentially constrain the
construction of a Type IV singularity. For loop quantum cosmology
\cite{new2} authors discussed the cases under which constrains of
baryon to entropy ratio well match with observations.

In the past few years, gravitational baryogenesis studied in various
modified theories of gravity. Some authors \cite{16,17} studied
baryogenesis phenomena in nonminimally coupled $f(R)$ theories and
$f(R)$ gravity respectively. They found only for tiny deviations of
a few percent, are consistent with the current bounds. In \cite{18},
Odintsov and Oikonomou investigated the ratio of the baryon number
to entropy density for the Gauss-Bonnet baryogenesis term while
Oikonomou and Saridakis \cite{19} discussed baryogenesis by
considering different cases of $f(T)$-gravity. Bento et al.
\cite{22} investigated baryogenesis in the framework of GB
braneworld cosmology, they also investigated the effect of the novel
terms on the baryon-to-entropy ratio. This mechanism were further
developed in minimal $f(R,\mathcal{T})$ gravity \cite{20} (where
$\mathcal{T}$ denotes the trace of stress energy momentum tensor) by
assuming that the universe is filled by dark energy and perfect
fluid. They explored cosmological gravitational baryogenesis
scenario through $f(R,\mathcal{T})=\alpha \mathcal{T}+\beta
\mathcal{T}^2+R$ and $f(R,\mathcal{T})=\lambda \mathcal{T}+R+\mu
R^2$ models (where $\alpha,~\beta,~\lambda$ and $\mu$ are non zero
coupling constants) and found constrains which are compatible with
the observation bounds. For non-minimal $f(R,\mathcal{T})$ gravity
\cite{21}, authors found that for terms proportional to
$\partial_\mu R$ and $\partial_\mu f(R,\mathcal{T})$ with suitable
parameter spaces, produced results that are consistent with
observations while interaction proportional to $\partial_\mu
\mathcal{T}$ produced unphysical result.

Moreover, Bhattacharjee and Sahoo \cite{23} explored baryogenesis in
$f(Q,\mathcal{T})$-gravity where $Q$ is the nonmetricity. They
considered $f(Q,\mathcal{T})=\alpha Q^{n+1}+\beta \mathcal{T}$ and
studied different baryogenesis interactions proportional to
$\partial_tQ$ and $(\partial_tQ) f_Q$, and found results that are
consistent with observations. Recently, Bhattacharjee \cite{23aaa}
worked on gravitational baryogenesis by using interactions
proportional to $\partial_iT$, $\partial_if(T)$, $\partial_i(T+B)$
and $\partial_if(T+B)$ and found excellent approximation for $f(T)$
and $f(T,B)$ theories of gravity. Whereas in case of $\partial_i(T +
B)$, author found unphysical results. In this work, we are
interested in investigating the gravitational baryogenesis mechanism
in the framework of $f(T,T_G)$- gravity as well as $f(T,B)$-gravity.
In the framework of $f(T, T_G)$-gravity we are taking two models
$f(T,T_G)= \alpha_1 \sqrt{T^2+\alpha_2 T_G}-T$ and
$f(T,T_G)=\alpha_1 T^2+\alpha_2 T \sqrt{|T_G|} + \beta_1
\sqrt{T^2+\beta_2 T_G}-T$, \cite{24} while for $f(T,B)$-gravity we
are considering $f(T,B)=-T+g(B)$ where $g(B)=f_1B \ln B$ and
$f(T,B)= A_0+A_1T+A_2T^2+A_3B^2+A_4TB$ (general Taylor expansion)
models. Arrangement of this paper as follow: In section \textbf{II},
we briefly introduce $f(T,T_G)$-gravity as well as $f(T,B)$-gravity.
Baryogenesis scenario for both theories of gravity discuss in
section \textbf{III}. Section \textbf{IV} is devoted to the study of
more complete and generalized baryogenesis interaction. Finally
conclusion are drawn in section \textbf{V}.

\section{Extended Teleparallel Theories of Gravity}

Here, we discuss the torsion based extended theories of gravity and
their field equations.

\subsection{$f(T,T_G)$-Gravity}

In this section, we briefly discuss some basic components of
teleparallel theory which leads to $f(T,T_G)$-gravity. Vierbein
fields $\left(e_A(x^\mu)\right)$ are the dynamical variables of
teleparallel gravity which can also expressed in components as
$e_a={e_a}^\mu\partial_\mu$. On the other hand, for dual vierbein,
it is defined as $e^a={e^a}_\mu dx^\mu$. The structure coefficients
arising from the vierbein commutation relation
$[e_a,e_b]=C^c_{ab}e_c,$ where $C^c_{ab}$ is defined as
\begin{eqnarray}\label{2}
C^c_{ab}=e_a^\mu e_b^\nu (e^c_{\mu,\nu}-e^c_{\nu,\mu}).
\end{eqnarray}
The torsion and curvature tensors in terms of tangent components are
given by
\begin{eqnarray}\label{3}
{T^a}_{bc}&=&{\omega^a}_{cb}-{\omega^a}_{bc}-{C^a}_{bc},\\\label{3+}
{R^a}_{bcd}&=&{\omega^a}_{bd,c}-{\omega^a}_{bc,d}+{\omega^e}_{bd}{\omega^a}_{ec}
-{\omega^e}_{bc}{\omega^a}_{ed}-{C^e}_{cd}{\omega^a}_{be},
\end{eqnarray}
where $\omega^a_{b}(x^\mu)$ is the connection $1$-form which defines
the source of parallel transformation. For an orthonormal vierbein,
the metric tensor is defined as $g_{\mu\nu}=\eta_{ab}e^a_\mu
e^b_\nu,$ where $\eta_{ab}=\textmd{diag}(-1,1,...1)$. Finally, it
proves convenient to define the torsion and contorsion tensors of
the form
\begin{eqnarray}\label{6}
{T^\lambda}_{\mu\nu}&=&{e_a}^\lambda(\partial_\nu
e^a_\mu-\partial_\mu
e^a_\nu),\\
{\mathcal{K}^{\mu\nu}}_\rho&=&-\frac{1}{2}({T^{\mu\nu}}_\rho-{T^{\nu\mu}}_\rho-{T_\rho}^{\mu\nu}).
\end{eqnarray}
Considering $R^a_{bcd}=0$ which is teleparallelism condition, one
can expresses the Weitzenb\"{o}ck connection as follows
$\bar{\omega}^\lambda_{\mu\nu}=e^\lambda_a e^a_{\mu,\nu}$. The Ricci
scalar in terms of usual Levi-Civita connection can be written as
$eR=-eT+2(eT^{\mu\nu}_\nu)_{,\mu}$ where
$e=\sqrt{|g|}=\textmd{det}({e^a}_\mu)$ and $T$ (torsion scalar) as
\begin{eqnarray}\label{10}
T=\frac{1}{4}T^{\mu\nu\lambda}T_{\mu\nu\lambda}+\frac{1}{2}T^{\mu\nu\lambda}T_{\lambda\nu\mu}-
{T^{\nu\mu}}_\nu {T_{\lambda\mu}}^\lambda.
\end{eqnarray}

The action defined by teleparallel gravity is
$S=\frac{1}{2\kappa^2}\int{eTd^4x}$ which is extended to the form
$S=\frac{1}{2\kappa^2}\int{e f(T)d^4x}$ as $f(T)$ theory action
\cite{nl1}-\cite{nl4}. Recently Kofinas, and Saridakis \cite{15d}
proposed teleparallel equivalent of Gauss-Bonnet (GB) theory by
coupling a new torsion scalar $T_G$, where the GB term $\bar{G}$ in
Levi-Civita connection is defined by
\begin{eqnarray}\label{12}
e\bar{G}=eT_G+ \texttt{total diverg}
\end{eqnarray}
where $T_G$ is defined as
\begin{eqnarray}\nonumber
T_G&=&\delta^{abcd}_{a_1a_2a_3a_4}({{\mathcal{K}}^{a_1}}_{ea}{{\mathcal{K}}^{ea_2}}_b
{{\mathcal{K}}^{a_3}}_{fc}{{\mathcal{K}}^{fa_4}}_d-
2{{\mathcal{K}}^{a_1a_2}}_a{{\mathcal{K}}^{a_3}}_{eb}{{\mathcal{K}}^e}_{fc}{{\mathcal{K}}^{fa_4}}_d+
2{{\mathcal{K}}^{a_1a_2}}_a\\\label{13}&\times&{{\mathcal{K}}^{a_3}}_{eb}{{\mathcal{K}}^{ea_4}}_f{{\mathcal{K}}^f}_{cd}
+2{{\mathcal{K}}^{a_1a_2}}_a{{\mathcal{K}}^{a_3}}_{eb}{{\mathcal{K}}^{ea_4}}_{c,d}).
\end{eqnarray}
where $\delta$ is the determinant of Kronecker deltas. The action
described by GB theory is $S=\frac{1}{2\kappa^2}\int{eT_Gd^4x}$. As
both theories $f(T)$ and $f(T_G)$ behave independently, so the
action involving both $T$ and $T_G$ is defined by
\begin{eqnarray}\label{15}
S=\frac{1}{2\kappa^2}\int{e f(T,T_G)d^4x},
\end{eqnarray}
which is clearly different from $f(T)$ and $f(R,G)$ theories of
gravity \cite{23c}. For $f(T,T_G)=-T$, it corresponds to
teleparallel gravity and one can obtained usual Einstein GB theory
for $F(T,T_G)=-T+\alpha T_G$, where $\alpha$ is the GB coupling.

In order to investigate the baryogenesis in $f(T,T_G)$-gravity, we
consider spatially flat FRW universe model as
\begin{eqnarray}\label{n1}
ds^2 = -dt^2 + a^2(t)dx_idx^i,
\end{eqnarray}
where $a(t)$ denotes the scale factor. This metric arises from the
diagonal vierbein ${e^a}_\mu =$diag$(1,a(t),a(t),a(t))$, so the
gravitational field equations for this geometry are given by
\begin{eqnarray}\label{tg1}
\rho_m&=&\frac{1}{2\kappa^2}(24H^3\dot{f}_{T_G}-12H^2f_T-T_Gf_{T_G}+f),\\\nonumber
p_m&=&-\frac{1}{2\kappa^2}\left(\frac{2}{3H}T_G\dot{f}_{T_G}
+8H^2\ddot{f}_{T_G}-4(3H^2+\dot{H})f_T-4H\dot{f}_T-T_Gf_{T_G}+f\right),\\\label{tg1111111}
\end{eqnarray}
where $p_m$ and $\rho_m$ are the pressure and energy density of
ordinary matter respectively, $H$ is the Hubble parameter such that
$H=\frac{1}{a(t)}\frac{d}{dt}a(t)$ and $f_T=\frac{\partial
f}{\partial T}$ , $f_{T_G}=\frac{\partial f}{\partial T_G}$, also
cosmic derivative of $f_{T_G}$ will be
$\dot{f}_{T_G}=f_{TT_G}\dot{T}+f_{{T_G}{T_G}}\dot{T}_G$. Finally
expressions for $T$ and $T_G$ read for FRW ansatz as
\begin{eqnarray}\label{tg3}
T&=&6H^2,\\T_G&=&24H^4+24H^2\dot{H}.
\end{eqnarray}
In case of $f(T,T_G)$-gravity, CP-violating interaction term of the
form,
\begin{eqnarray}\label{tg5}
\frac{1}{M_*^2}\int{\sqrt{-g}(\partial_\mu(T+T_G))J^\mu dx^4}.
\end{eqnarray}
In this case, baryon to entropy ratio can be defined as
\begin{eqnarray}\label{tg6}
\frac{\eta_B}{s}\simeq -\frac{15g_b}{4\pi^2
g_*}\left(\frac{\dot{T}+\dot{T}_G}{M_*^2\mathcal{T}}\right)|_{\mathcal{T}_D},
\end{eqnarray}
where $\mathcal{T}_D$, denotes the decoupling temperature while
$g_b$ and $g_*$ are the total number of intrinsic degrees of freedom
of baryon and number of the degrees of freedom of the effectively
massless particles.  In this paper we assume the existence of
thermal equilibrium which prevails with energy density being
associated with temperature $\mathcal{T}_D$ as,
\begin{eqnarray}\label{tg7}
\rho=\frac{\pi^2}{30}g_*\mathcal{T}_D^4.
\end{eqnarray}

In the framework of $f(T,T_G)$-gravity, we focus on two particular
models which are:
\begin{itemize}
\item \textbf{Model 1:} $f(T,T_G)= \alpha_1\sqrt{T^2+\alpha_2T_G}-T$
\item \textbf{Model 2:} $f(T,T_G)
=\alpha_1T^2+\alpha_2T\sqrt{\mid
T_G\mid}+\beta_1\sqrt{T^2+\beta_2T_G}-T$
\end{itemize}
where all $\alpha_i$ and $\beta_i$ are dimensionless coupling
parameters. These models contain some torsion based terms which make
these models as generalizations of $f(T)$ gravity. Since
teleparallel gravity inherits linear torsion term while $f(T)$
generalized the torsion scalar by adding its quadratic form which is
the most simple $f(T)$ model. In the similar way, as in Model I, $T$
and $\sqrt{T^2+\alpha_2T_G}$ have the same order because $T_G$ keeps
quartic power of torsion scalar. This model is said to be simplest
and non-trivial due to same order of terms which results no extra
mass scale in the modification of theory and also modified the
teleparallel gravity. Taking $\alpha_1=0$ or $\alpha_2=0$ lead to
teleparallel gravity or equivalently, to general relativity. The
Model I is expected to discuss the late-time cosmological scenarios.
We restrict our discussions for the case $\alpha_2\neq0$.

In order to discuss early times of cosmic expansion, Model I needs
to modify by introducing higher order terms like $T^2$. As $T_G$ is
of same order with quadric torsion scalar, so it must be included in
the action of model framework. However, it is not included as it is
because $T_G$ is topological in four dimensions. Thus, the term
$T\sqrt{\mid T_G\mid}$ which is also of the same order with $T^2$
and nontrivial added in the action. Thus, this form of unified
action develops a gravitational theory which gives the description
about inflation as well as late cosmic expansion of the universe
with acceleration. Initially, these models were used in \cite{15d},
in which authors investigated the phase space analysis and expansion
history from early-times to late-times cosmic acceleration and found
that the effective equation of state parameter can represents
different eras of the universe namely, quintessence, phantom and
quintom phase. Also, Minkowski stability problem in
$f(T,T_G)$-gravity was discussed by considering these models
\cite{mot1}.

\subsection{$f(T,B)$ Gravity}

Recently, Bahamonde et al. \cite{15e} constructed a new modification
of standard $f(T)$-gravity by involving a boundary term $B$ with
$R$. The action in $f(T,B)$ is given as
\begin{eqnarray}\label{a2}
S=\frac{1}{2\kappa^2}\int{e f(T,B)d^4x}.
\end{eqnarray}
In \cite{15e} it was proposed that for $f(T,B)=f(T)$ and
$f(T,B)=f(-T+B)=f(R)$, one can recover both $f(T)$ and $f(R)$
gravity theories, respectively. Varying action in Eq. (\ref{a2})
with respect to the tetrad field, we get the field equations
\begin{eqnarray}\nonumber
16\pi e\mathcal{T}^\lambda_\nu+ef\delta^\lambda_\nu &=&
\left(2e\delta^\lambda_\nu\Box -2e\nabla^\lambda \nabla_\nu + eB
\delta^\lambda_\nu\right)f_B +4e\left[(\partial_\mu
f_B)+(\partial_\mu
f_T)\right]S_\nu^{\mu\lambda}+4e^a_\nu\\\label{a3}&\times&
\partial_\mu (eS_a^{\mu\lambda})f_T-4ef_T
T^\sigma_{\mu\nu}S_\sigma^{\lambda\mu},
\end{eqnarray}
where $f_B=\frac{\partial f}{\partial B}$, $\Box=\nabla^\mu
\nabla_\mu$. Evaluating  Eq.(\ref{a3}), Friedmann equations turn out
to be \cite{27}-\cite{29}
\begin{eqnarray}\label{tb7}
-3H^2(3f_B+2f_T)+3H\dot{f}_B-3\dot{H}f_B+\frac{1}{2}f&=&\kappa^2\rho_m,\\\label{tb7777777}
-(3H^2+\dot{H})(3f_B+2f_T)-2H\dot{f}_T+\ddot{f}_B+\frac{1}{2}f&=&-\kappa^2p_m.
\end{eqnarray}
where the expressions for $T$ and $B$ are
\begin{eqnarray}\label{tb3}
T=6H^2,~~~~B=18H^2+6\dot{H}.
\end{eqnarray}
Together these form the Ricci scalar as $R=-T+B=12H^2+6\dot{H}$.
This shows how $f(R)$ gravity results as a subset of
$f(T,B)$-gravity where $f(T,B):=f(-T+B)=f(R)$. For $f(T,B)$-gravity,
CP-violating term is given in the form,
\begin{eqnarray}\label{tb9}
\frac{1}{M_*^2}\int{\sqrt{-g}(\partial_\mu(T+B))J^\mu dx^4}.
\end{eqnarray}
The baryon to entropy ratio for $f(T,B)$-gravity becomes
\begin{eqnarray}\label{tb10}
\frac{\eta_B}{s}\simeq -\frac{15g_b}{4\pi^2
g_*}\left(\frac{\dot{T}+\dot{B}}{M_*^2\mathcal{T}}\right)|_{\mathcal{T}_D}.
\end{eqnarray}

We focus our attention on two particular $f(T,B)$ models
(logarithmic and general Taylor expansion model), which are:
\textbf{\begin{itemize}
  \item \textbf{Model III:} $f(T,B)=-T+g(B)$, where $g(B)=f_1B \ln
  B$,
  \item \textbf{Model IV:} $f(T,B)=A_0+A_1T+A_2T^2+A_3B^2+A_4TB$,
\end{itemize}}
where $A_i$ are numerical constants. These models are modified
models where the logarithmic as well as quadratic and product
boundary terms are added to contribute in modification of
teleparallel gravity. In \cite{mot2}, authors demonstrated that the
behavior of these models can undergo an epoch of late-time
acceleration and reproduced quintessence and phantom regimes with a
transition along the phantom-divided line. Same authors \cite{mot3}
studied cosmological solution of the $f(T,B)$-gravity, using
dynamical system analysis against model IV and found constrains
which favor current observational data.

\section{Baryogenesis}

Here, we investigate the baryogenesis of above listed models of
$f(T,T_G)$ (Models I and II) and $f(T,B)$ (Models III and IV)
theories of gravity. We consider power-law form of scale factor as
$a(t) = m_0t^\gamma$, (where $m_0$ and $\gamma$ are the non zero
parameter) for each model and construct baryon to entropy ratio.

\subsection{Model I}

For this model, we develop baryon to entropy ratio in terms of
decoupling temperature $\mathcal{T}_D$. So for this purpose, we find
energy density $\rho$ in terms of decoupling cosmic time $t_D$.
Initially, we find the corresponding expressions $f_T$, $f_{T_G}$
and $\dot{f}_{T_G}$, which can be calculated as
\begin{eqnarray}\label{tg8}
f_T&=&-1+\frac{6\alpha_1}{\sqrt{36+24\alpha_2-\frac{24\alpha_2}{\gamma}}},\\
f_{T_G}&=&\frac{\alpha_1\alpha_2t^2}{2\gamma^2\sqrt{36+24\alpha_2-\frac{24\alpha_2}{\gamma}}},\\
\dot{f}_{T_G}&=&\frac{36\alpha_1\alpha_2t}{\gamma^2\left(36+24\alpha_2-\frac{24\alpha_2}{\gamma}\right)^\frac{3}{2}}
+\frac{24\alpha_1\alpha_2^2(\gamma-1)t}
{\gamma^3\left({36+24\alpha_2-\frac{24\alpha_2}{\gamma}}\right)^\frac{3}{2}}.
\end{eqnarray}
Inserting these equations in (\ref{tg1}), we obtain the energy
density as follows
\begin{eqnarray}\nonumber
\rho&=&\frac{1}{2\kappa^2t^2}\bigg(6\gamma^2+\alpha_1\gamma^2A-\frac{72\alpha_1\gamma^2}{A}-
\frac{12\alpha_1\alpha_2\gamma(\gamma-1)}{A}+\frac{864\alpha_1\alpha_2\gamma}{A^3}
\\\label{tg11}&+&\frac{576\alpha_1\alpha_2^2(\gamma-1)}{A^3}\bigg),
\end{eqnarray}
where $A=\sqrt{36+24\alpha_2-\frac{24\alpha_2}{\gamma}}$. Equating
Eqs. (\ref{tg7}) and (\ref{tg11}), we obtain $t_D$ as a function of
$\mathcal{T}_D$ is given by
\begin{eqnarray}\nonumber
t_D&=&\frac{1}{\kappa \pi
\mathcal{T}_D^2}\bigg(\frac{15}{g_*}\bigg(6\gamma^2+\alpha_1\gamma^2A-\frac{72\alpha_1\gamma^2}{A}-
\frac{12\alpha_1\alpha_2\gamma(\gamma-1)}{A}+\frac{864\alpha_1\alpha_2\gamma}{A^3}
\\\label{tg12}&+&\frac{576\alpha_1\alpha_2^2(\gamma-1)}{A^3}\bigg)\bigg)^\frac{1}{2}.
\end{eqnarray}
Thus the expression of net baryon to entropy ratio for this specific
model can be obtained by using Eqs. (\ref{tg6}) and (\ref{tg12}) as
follows
\begin{eqnarray}\nonumber
\frac{\eta_B}{S}&\simeq& \frac{45g_b\gamma^2 \kappa^3 \pi
\mathcal{T}_D^5}{g_*
M_*^2}\bigg(\frac{15}{g_*}\bigg(6\gamma^2+\alpha_1\gamma^2A-\frac{72\alpha_1\gamma^2}{A}-
\frac{12\alpha_1\alpha_2\gamma(\gamma-1)}{A}\\\nonumber&+&\frac{864\alpha_1\alpha_2\gamma}{A^3}
+\frac{576\alpha_1\alpha_2^2(\gamma-1)}{A^3}\bigg)\bigg)^{-\frac{3}{2}}
\bigg(1+8\pi^2\kappa^2\mathcal{T}_D^4\gamma(\gamma-1)\bigg(\frac{15}{
 g_*}\\\nonumber&\times&
\bigg(6\gamma^2+\alpha_1\gamma^2A-\frac{72\alpha_1\gamma^2}{A}-
\frac{12\alpha_1\alpha_2\gamma(\gamma-1)}{A}+\frac{864\alpha_1\alpha_2\gamma}{A^3}
\\\label{tg13}&+&\frac{576\alpha_1\alpha_2^2(\gamma-1)}{A^3}\bigg)\bigg)^{-1}\bigg).
\end{eqnarray}
\begin{figure}
\centering \epsfig{file=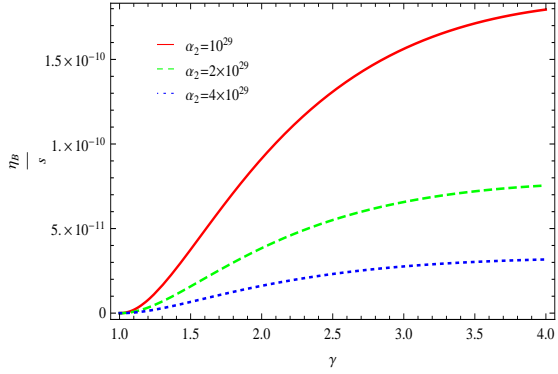, width=.45\linewidth,
height=2.1in} \caption{Plot of baryon to entropy ratio
$\frac{\eta_B}{S}$ versus $\gamma$ for Model I for different values
of $\alpha_2$, other parameters are $g_b=1,~
\mathcal{T}_D=2\times10^{16},~M_\ast=10^{12},~g_\ast=106,~ \kappa=1$
and $\alpha_1=2\times10^{39}$.}
\end{figure}
In Figure \textbf{1}, we plot baryon to entropy ratio in terms of
parameter $\gamma$ for different values of $\alpha_2$. For
$\alpha_2=10^{29}$ and $\gamma=2$, we can see baryon to entropy
ratio is confined to $\frac{\eta_B}{S}=8.9\times10^{-11}$, also
showing compatibility with observations. For other values of
$\alpha_2$, we obtain results which are compatible with the
observational value. Following Table \textbf{1} shows the different
approach of baryon to entropy ratio for $\gamma=2,~3,~4$.
\begin{table}[ht]
\caption{Baryogenesis for
$f(T,T_G)=\alpha_1\sqrt{T^2+\alpha_2T_G}-T$} \centering
\begin{tabular}{c c c}
\hline\hline
$\alpha_2$ ~~~~~~~~~~~& $\gamma$~~~~~~~~~~~ & $\frac{\eta_B}{S}$ (Baryon to entropy ratio) \\
[0.5ex] \hline
$10^{29}$  ~~~~~~~~~~~& $2$~~~~~~~~~~~ & $8.9\times10^{-11}$ \\
$2\times10^{29}$  ~~~~~~~~~~~& $3$~~~~~~~~~~~ & $6.5\times10^{-11}$ \\
$4\times10^{29}$  ~~~~~~~~~~~& $4$~~~~~~~~~~~ & $3.1\times10^{-11}$ \\[1ex]
\hline
\end{tabular}
\label{table:nonlin}
\end{table}

\subsection{Model II}

This model is obtained from previous model by adding higher order
correction terms $T^2$ and $T\sqrt{\mid T_G\mid}$. For this model,
we also find the expressions $f_T$, $f_{T_G}$ and $\dot{f}_{T_G}$,
which are obtained as
\begin{eqnarray}\label{tg14}
f_T&=&-1+\frac{12\alpha_1\gamma^2}{t^2}+
\frac{6\beta_1}{\sqrt{36+24\beta_2-\frac{24\beta_2}{\gamma}}}+
\frac{\alpha_2\gamma^2\sqrt{|24-\frac{24}{\gamma}|}}{t^2},\\
f_{T_G}&=&\frac{\beta_1\beta_2t^2}{2\gamma^2\sqrt{36+24\beta_2-\frac{24\beta_2}{\gamma}}}
+\frac{3\alpha_2}{\sqrt{|24-\frac{24}{\gamma}|}},\\
\dot{f}_{T_G}&=&\frac{36\beta_1\beta_2t}{\gamma^2\bigg(36+24\beta_2-\frac{24\beta_2}{\gamma}\bigg)^{\frac{3}{2}}}
+\frac{24\beta_1\beta_2^2(1-\frac{1}{\gamma})t}
{\gamma^2\bigg(36+24\beta_2-\frac{24\beta_2}{\gamma}\bigg)^{\frac{3}{2}}}.
\end{eqnarray}
Substituting these expressions in Eq. (\ref{tg1}), we have
\begin{eqnarray}\nonumber
\rho&=&\frac{1}{2\kappa^2
t^2}\bigg(6\gamma^2+\beta_1\gamma^2A-\frac{72\beta_1\gamma^2}{A}-\frac{\beta_1\beta_2\gamma^4\zeta^2}{2A}+
\frac{864\beta_1\beta_2\gamma}{A^3}+\frac{24\beta_1\beta_2^2\gamma
\zeta^2}{A^3}\bigg)\\\label{tg17}&-&\frac{1}{2\kappa^2t^4}\bigg(108\alpha_1\gamma^4+9\alpha_2\gamma^4\zeta\bigg),
\end{eqnarray}
where $\zeta=\sqrt{|24-\frac{24}{\gamma}|}$. Comparing Eqs.
(\ref{tg7}) with (\ref{tg17}), we obtain $t_D$ as
\begin{eqnarray}\label{tg18}
t_D=\left(\frac{c_1+\sqrt{c_1^2-\frac{4\kappa^2\pi^2g_\ast
c_2\mathcal{T}_D^4}{15}}}
{\frac{2\kappa^2\pi^2g_\ast\mathcal{T}_D^4}{15}}\right)^{\frac{1}{2}},
\end{eqnarray}
where
$c_1=6\gamma^2+\beta_1\gamma^2A-\frac{72\beta_1\gamma^2}{A}-\frac{\beta_1\beta_2\gamma^4\zeta^2}{2A}+
\frac{864\beta_1\beta_2\gamma}{A^3}+\frac{24\beta_1\beta_2^2\gamma
\zeta^2}{A^3}$ and $c_2=108\alpha_1\gamma^4+9\alpha_2\gamma^4\zeta$.
Using Eq.(\ref{tg18}), we obtain the final expression for this
particular model as
\begin{eqnarray}\nonumber
\frac{\eta_B}{S}&\simeq& \frac{45g_b\gamma^2 \kappa^3 \pi
\mathcal{T}_D^5}{g_* M_*^2}\bigg(1+\frac{16\kappa^2\pi^2g_\ast
\mathcal{T}_D^4
\gamma(\gamma-1)}{15\bigg(c_1+\sqrt{c_1^2-\frac{4\kappa^2\pi^2g_\ast
c_2
\mathcal{T}_D^4}{15}}\bigg)}\bigg)\bigg(\sqrt{c_1^2-\frac{4\kappa^2\pi^2g_\ast
c_2
\mathcal{T}_D^4}{15}}+c_1\bigg)^{-\frac{3}{2}}\\\label{tg19}&\times&\bigg(\frac{2g_\ast}{15}\bigg)^{\frac{3}{2}}.
\end{eqnarray}
\begin{figure}
\centering \epsfig{file=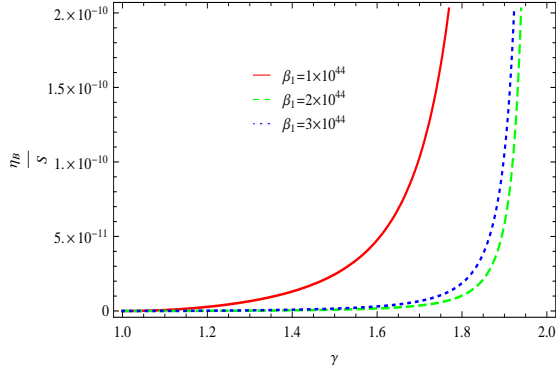, width=.45\linewidth,
height=2.1in} \caption{The behavior of baryon to entropy ratio
$\frac{\eta_B}{S}$ versus $\gamma$ for Model II, for $g_b=1,~
\mathcal{T}_D=2\times10^{16},~M_\ast=10^{12},~g_\ast=106,~
\kappa=1,~\alpha_1=2\times10^{-20},~\alpha_2=2\times10^{20}$ and
$\beta_2=10^{20}$.}
\end{figure}

Figure \textbf{2} illustrates the dependence of the baryon to
entropy ratio on the dimensionless parameter $\gamma$ for
\textbf{Model II}. We notice that when $1.65\leq\gamma\leq 1.94$, we
obtain $\frac{\eta_B}{S}$ in leading order as
$7.5^{+1.5}_{-1.1}\times 10^{-11}$ which is compatible with
observational bounds. Following table describes the detailed
discussion of Figure \textbf{2}.

\begin{table}[]
\caption{Baryogenesis for $f(T,T_G) =\alpha_1T^2+\alpha_2T\sqrt{\mid
T_G\mid}+\beta_1\sqrt{T^2+\beta_2T_G}-T$} \centering
\begin{tabular}{c c c}
\hline\hline
$\beta_1$ ~~~~~~~~~~~& $\gamma$~~~~~~~~~~~ & $\frac{\eta_B}{S}$ (Baryon to entropy ratio) \\
[0.5ex] \hline
$10^{44}$  ~~~~~~~~~~~& $1.65$~~~~~~~~~~~ & $6.9\times10^{-11}$ \\
$2\times10^{44}$  ~~~~~~~~~~~& $1.9$~~~~~~~~~~~ & $7\times10^{-11}$ \\
$3\times10^{44}$  ~~~~~~~~~~~& $1.87$~~~~~~~~~~~ & $7.2\times10^{-11}$ \\[1ex]
\hline
\end{tabular}
\label{table:nonlin}
\end{table}

\subsection{Model III}

Bahamonde and Capozziello \cite{26} investigated this model by
considering $g(B)=f_1B\ln B$ where $f_1$ is an arbitrary constant.
So expressions $f_T,~f_B$ and $\dot{f}_B$, for this model will be as
follows
\begin{eqnarray}\label{tb11}
f_T=-1,\quad
f_B=f_1\left(1+\ln\left(\frac{6\gamma(3\gamma-1)}{t^2}\right)\right),
\quad\dot{f}_B=\frac{-2f_1}{t}.
\end{eqnarray}
Now, one can find the energy density of ordinary matter $\rho(t)$ by
using Eqs. (\ref{tb7}) and (\ref{tb11})
\begin{eqnarray}\label{tb12}
\rho(t)=\frac{1}{\kappa^2t^2}\left(3\gamma^2-3\gamma f_1-9\gamma^2
f_1\right).
\end{eqnarray}
Using Eqs.(\ref{tg7}) and (\ref{tb12}), we get $t_D$ as
\begin{eqnarray}\label{tb13}
t_D=\frac{3\sqrt{10}}{\kappa\pi
\mathcal{T}_D^2}\sqrt{\frac{\gamma\left(\gamma-3f_1\gamma-f_1\right)}{g_\ast}}.
\end{eqnarray}
Now expression of baryon to entropy ratio can be obtained by using
Eqs. (\ref{tb7}), (\ref{tb3}), (\ref{tb10}) and (\ref{tb13}) as
follow
\begin{eqnarray}\label{tb14}
\frac{\eta_B}{S}\simeq \frac{\kappa^3\pi
\mathcal{T}_D^5(4\gamma-1)g_b\sqrt{g_\ast}}{6\sqrt{10}M_\ast^2\gamma^{\frac{1}{2}}
\left(\gamma-3f_1\gamma-f_1\right)^{\frac{3}{2}}}.
\end{eqnarray}
\begin{figure}
\centering \epsfig{file=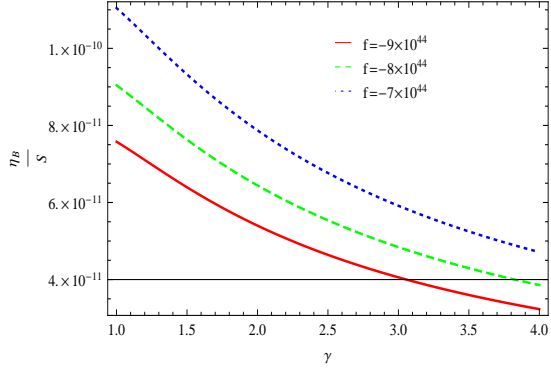, width=.45\linewidth, height=2.1in}
\caption{Plot of $\frac{\eta_B}{S}$ as the function of $\gamma$ for
Model III, we take
$g_b=1,~\mathcal{T}_D=2\times10^{16},~M_\ast=10^{12},~g_\ast=106,~\kappa=1$.}
\end{figure}
In Figure \textbf{3}, we plot the baryon to entropy ratio against
parameter $\gamma$. As it can be seen when $\gamma\leq1.56$, baryon
to entropy ratio lies in the range
$7.5^{+1.5}_{-1.5}\times10^{-11}$, which favors the observational
value. \textbf{Table III} indicates the different cases of baryon to
entropy ratio.
\begin{table}[]
\caption{Baryogenesis for $f(T,B)=-T+g(B)$} \centering
\begin{tabular}{c c c}
\hline\hline
$f$ ~~~~~~~~~~~& $\gamma$~~~~~~~~~~~ & $\frac{\eta_B}{S}$ (Baryon to entropy ratio) \\
[0.5ex] \hline
$-9\times10^{44}$  ~~~~~~~~~~~& $1$~~~~~~~~~~~ & $7.5\times10^{-11}$ \\
$-8\times10^{44}$  ~~~~~~~~~~~& $1$~~~~~~~~~~~ & $9\times10^{-11}$ \\
$-7\times10^{44}$  ~~~~~~~~~~~& $1.5$~~~~~~~~~~~ & $9.3\times10^{-11}$ \\[1ex]
\hline
\end{tabular}
\label{table:nonlin}
\end{table}

\subsection{Model IV}

First we consider a general Taylor expansion of the $f(T,B)$
Lagrangian \cite{25a} as
\begin{eqnarray}\nonumber
f(T,B)&=&f(T_0,B_0)+f_T(T_0,B_0)(T-T_0)+f_B(T_0,B_0)(B-B_0)+\frac{1}{2!}f_{TT}\\\nonumber&\times&(T_0,B_0)(T-T_0)^2
+\frac{1}{2!}f_{BB}(T_0,B_0)(B-B_0)^2+f_{TB}(T_0,B_0)(T\\\label{b1}&-&T_0)(B-B_0)+\mathcal{O}(T^3,B^3),
\end{eqnarray}
Since boundary term $B$ has linear order, so consider $T_0=B_0=0$,
by taking constants $A_i$, the Lagrangian can be written as
\begin{eqnarray}\label{b2}
f(T,B) =A_0+A_1T+A_2T^2+A_3B^2+A_4TB.
\end{eqnarray}
Next, we find the expressions $f_T$, $f_B$ and $\dot{f}_B$, which
lead to
\begin{eqnarray}\label{tb15}
f_T&=&A_1+\frac{12A_2\gamma^2}{t^2}+\frac{6A_4\gamma(3\gamma-1)}{t^2},\\\label{tb16}
f_B&=&\frac{12A_3\gamma(3\gamma-1)}{t^2}+\frac{6A_4\gamma^2}{t^2},\\\label{tb17}
\dot{f}_B&=&-\frac{24A_3\gamma(3\gamma-1)}{t^3}-\frac{12A_4\gamma^2}{t^3}.
\end{eqnarray}
Using Eqs. (\ref{tb7}), (\ref{tb15}), (\ref{tb16}) and (\ref{tb17}),
on can write the energy density $\rho(t)$ in a radiation dominated
universe as
\begin{eqnarray}\label{tb18}
\rho(t)
&=&\frac{1}{\kappa^2t^4}\left(-162A_3\gamma^4-108A_3\gamma^3-108A_4\gamma^4-54A_2\gamma^4+54A_3\gamma^2\right)-
\frac{3A_1\gamma^2}{\kappa^2t^2}+\frac{A_0}{2\kappa^2}.
\end{eqnarray}
Decoupling cosmic time for this case, will be
\begin{eqnarray}\label{tb19}
t_D=\left(\frac{2\chi}{3A_1\gamma^2+\sqrt{9A_1^2\gamma^4+2\chi\left(\frac{\kappa^2
\pi^2g_\ast \mathcal{T}_D^4}{15}-A_0\right)}}\right)^{\frac{1}{2}},
\end{eqnarray}
where
$\chi=-162A_3\gamma^4-108A_3\gamma^3-108A_4\gamma^4-54A_2\gamma^4+54A_3\gamma^2$.
In this case, baryon to entropy ration will be
\begin{eqnarray}\label{tb20}
\frac{\eta_B}{S}=\frac{45g_b\gamma(4\gamma-1)}{\pi^2g_\ast M_\ast^2
\mathcal{T}_D}\left(\frac{3A_1\gamma^2+\sqrt{9A_1^2\gamma^4+2\chi\left(\frac{\kappa^2
\pi^2g_\ast
\mathcal{T}_D^4}{15}-A_0\right)}}{2\chi}\right)^{\frac{3}{2}}.
\end{eqnarray}
\begin{figure}
\centering \epsfig{file=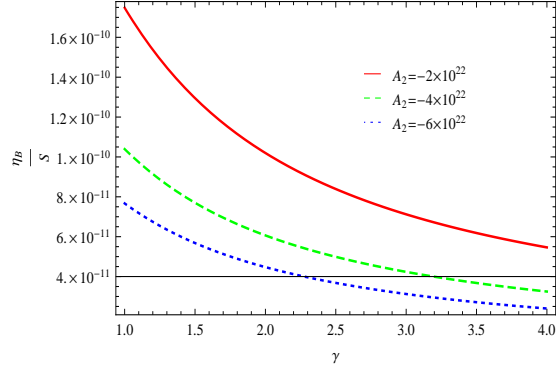, width=.45\linewidth, height=2.1in}
\caption{Plot of baryon to entropy ratio $\frac{\eta_B}{S}$ against
$\gamma$ for Model IV, for $g_b=1,~\mathcal{T}_D=2\times10^{16},~
M_\ast=10^{12},~g_\ast=106,~\kappa=1,~A_0=2\times10^{10},~
A_1=3\times10^{10},~A_3=5\times10^{10}$ and $A_4=6\times10^{10}$.}
\end{figure}
Figure \textbf{4} yields the baryon to entropy ratio verses $\gamma$
in the framework of $f(T,B)$-gravity with general Taylor expansion
model for different values of $A_2$. One can see that for
$A_2=-2\times10^{22}$, before $\gamma=2.5$, baryon to entropy ratio
is $5.5\times10^{-11}\leq\frac{\eta_B}{S}\leq8.09\times10^{-11}$.
Moreover, for other cases when $\gamma\geq1.25$, the trajectories
are ruled out by observationally measured value of
$\frac{\eta_B}{S}$. \textbf{Table IV} also summarizes some values of
baryon to entropy ratio for $\gamma=1,~1.1,~2$
\begin{table}[ht]
\caption{Baryogenesis for $f(T,B) =A_0+A_1T+A_2T^2+A_3B^2+A_4TB$}
\centering
\begin{tabular}{c c c}
\hline\hline
$A_2$ ~~~~~~~~~~~& $\gamma$~~~~~~~~~~~ & $\frac{\eta_B}{S}$ (Baryon to entropy ratio) \\
[0.5ex] \hline
$-2\times10^{22}$  ~~~~~~~~~~~& $2$~~~~~~~~~~~ & $9\times10^{-11}$ \\
$-4\times10^{22}$  ~~~~~~~~~~~& $1.1$~~~~~~~~~~~ & $9\times10^{-11}$ \\
$-6\times10^{22}$  ~~~~~~~~~~~& $1$~~~~~~~~~~~ & $7.99\times10^{-11}$ \\[1ex]
\hline
\end{tabular}
\label{table:nonlin}
\end{table}

\section{Generalized Baryogenesis Interaction}

In this section, we present the more complete and generalized
baryogenesis interaction  in the framework of $f(T,T_G)$-gravity
\cite{23aaa,19}. For this case CP-violation interaction proportional
to $\partial_\mu f(T+T_G)$, can be written as
\begin{eqnarray}\label{ali1}
\frac{1}{M_\ast}\int{\sqrt{-g}d^4x(\partial_\mu f(T+T_G))J^\mu}.
\end{eqnarray}
For this kind of baryogenesis interaction, baryon to entropy ratio
will be as follows
\begin{eqnarray}\label{ali2}
\frac{\eta_B}{S}\simeq -\frac{15g_b}{4\pi^2
g_*}\left(\frac{\dot{T}f_T+\dot{T}_Gf_{T_G}}{M_*^2\mathcal{T}}\right)|_{\mathcal{T}_D}.
\end{eqnarray}
For this case CP-violation interaction term in the framework of
$f(T,B)$-gravity written as
\begin{eqnarray}\label{z1}
\frac{1}{M_\ast}\int{\sqrt{-g}d^4x(\partial_\mu f(T+B))J^\mu}.
\end{eqnarray}
Using Eq. (\ref{z1}), baryon to entropy ratio is given by
\begin{eqnarray}\label{z2}
\frac{\eta_B}{s}\simeq -\frac{15g_b}{4\pi^2
g_*}\left(\frac{\dot{T}f_T+\dot{B}f_{B}}{M_*^2\mathcal{T}}\right)|_{\mathcal{T}_D}.
\end{eqnarray}

\subsection{Model I}

Using Eqs. (\ref{tg12}) and (\ref{ali2}), we have the following
expression of baryon to entropy ratio
\begin{eqnarray}\nonumber
\frac{\eta_B}{S} &=& \frac{45g_b\kappa^3\pi
\mathcal{T}_D^5\gamma}{g_\ast
M_\ast^2}\left(\frac{6\gamma\alpha_1}{A}+\frac{4(\gamma-1)\alpha_1\alpha_2}{A}-\gamma\right)
\bigg(\frac{15}{g_\ast}\bigg(6\gamma^2+
A\gamma^2\alpha_1\\\nonumber&-&\frac{72\gamma^2\alpha_1}{A}-\frac{12\gamma(\gamma-1)\alpha_1\alpha_2}{A}
\frac{864\gamma\alpha_1\alpha_2}{A^3}+
\frac{576\alpha_1\alpha_2^2(\gamma-1)}{A^3}\bigg)\bigg)^{-\frac{3}{2}}.\\\label{ali3}&&
\end{eqnarray}
\begin{figure}
\centering \epsfig{file=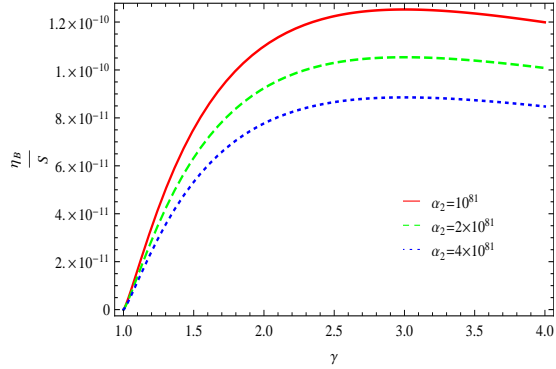, width=.45\linewidth,
height=2.1in} \caption{Plot of baryon to entropy ratio
$\frac{\eta_B}{S}$ versus $\gamma$ for generalized baryogenesis
interaction for Model I for different values of $\alpha_2$, with
$g_b=1,~\mathcal{T}_D=2\times10^{16},~M_\ast=10^{12},~g_\ast=106,~
\kappa=1$ and $\alpha_1=10^{94}$.}
\end{figure}
In case of generalized baryogenesis interaction, the graph of baryon
to entropy ratio verses $\gamma$ parameter is shown in Figure
\textbf{5} for different values of $\alpha_2$. Thus three different
cases can be distinguished as
\begin{itemize}
  \item For $\alpha_2=10^{81}$ and $1.15\lesssim \gamma\lesssim 1.5$, we have
   $2\times10^{-11}\lesssim  \frac{\eta_B}{S}\lesssim
  7.5\times10^{-11}$.
  \item For $\alpha_2=2\times10^{81}$ and $1.15\lesssim \gamma\lesssim
  2$, then baryon to entropy ratio lies in the range $2\times10^{-11}\lesssim  \frac{\eta_B}{S}\lesssim
 9.4\times10^{-11}$.
  \item For $\alpha_2=4\times10^{81}$ and $1.15\lesssim \gamma\lesssim
  2.5$, we have $2\times10^{-11}\lesssim  \frac{\eta_B}{S}\lesssim
 8.6\times10^{-11}$.
\end{itemize}
All constraints are very close to the observationally accepted
value. Other cases of baryon to entropy ratio are discussed in
\textbf{Table V}.
\begin{table}[]
\caption{Generalized Baryogenesis Interaction for
$f(T,T_G)=\alpha_1\sqrt{T^2+\alpha_2T_G}-T$} \centering
\begin{tabular}{c c c}
\hline\hline
$\alpha_2$ ~~~~~~~~~~~& $\gamma$~~~~~~~~~~~ & $\frac{\eta_B}{S}$ (Baryon to entropy ratio) \\
[0.5ex] \hline
$10^{81}$  ~~~~~~~~~~~& $1.5$~~~~~~~~~~~ & $7.5\times10^{-11}$ \\
$2\times10^{81}$  ~~~~~~~~~~~& $2$~~~~~~~~~~~ & $9.4\times10^{-11}$ \\
$4\times10^{81}$  ~~~~~~~~~~~& $2.5$~~~~~~~~~~~ & $8.6\times10^{-11}$ \\[1ex]
\hline
\end{tabular}
\label{table:nonlin}
\end{table}

\subsection{Model II}

For generalized baryogenesis interaction case, the baryon to entropy
ratio (\ref{ali2}) for this specific model become
\begin{eqnarray}\nonumber
\frac{\eta_B}{S}&=&\frac{45g_b\gamma^2 \kappa^3 \pi
\mathcal{T}_D^5}{g_*
M_*^2}\bigg(\frac{2g_\ast}{15}\bigg)^{\frac{3}{2}}\bigg(c_1+\sqrt{c_1^2-\frac{4\kappa^2\pi^2g_\ast
c_2
\mathcal{T}_D^4}{15}}\bigg)^{-\frac{3}{2}}\bigg(\frac{6\beta_1}{A}\\\label{alii1}&+&\frac{2\beta_1\beta_2(\gamma-1)}{A\gamma}+
\frac{\left(2\kappa^2\pi^2g_\ast
\mathcal{T}_D^4\right)\left(12\alpha_1\gamma^2+\alpha_2\gamma^2\zeta+\frac{12\alpha_2\gamma(\gamma-1)}
{\zeta}\right)}{15\left(c_1+\sqrt{c_1^2-\frac{4\kappa^2 \pi^2g_\ast
\mathcal{T}_D^4c_2}{15}}\right)}\bigg).
\end{eqnarray}
\begin{figure}
\centering \epsfig{file=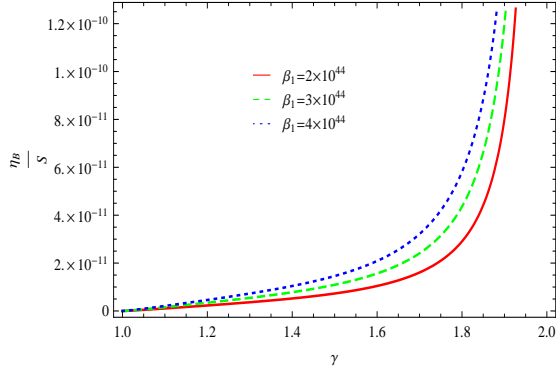, width=.45\linewidth,
height=2.1in} \caption{Plot of baryon to entropy ratio
$\frac{\eta_B}{S}$ against $\gamma$ in the context of generalized
baryogenesis interaction for Model II, in this case
$g_b=1,~\mathcal{T}_D=2\times10^{16},~M_\ast=10^{12},~g_\ast=106,~
\kappa=1,~\alpha_1=2\times10^{40},~\alpha_2=2\times10^{70}$ and
$\beta_2=10^{90}$.}
\end{figure}
Graphical behavior of Eq. (\ref{alii1}) is shown in Figure
\textbf{6} for different values of $\beta_1$, one can notice all
trajectories are correspond to $\frac{\eta_B}{S}=7.9\times10^{-11}$
when $\gamma=1.9$, $\gamma=1.85$ and $\gamma=1.83$ as mention in
following \textbf{Table VI}.
\begin{table}[]
\caption{Generalized Baryogenesis Interaction for $f(T,T_G)
=\alpha_1T^2+\alpha_2T\sqrt{\mid
T_G\mid}+\beta_1\sqrt{T^2+\beta_2T_G}-T$} \centering
\begin{tabular}{c c c}
\hline\hline
$\beta_1$ ~~~~~~~~~~~& $\gamma$~~~~~~~~~~~ & $\frac{\eta_B}{S}$ (Baryon to entropy ratio) \\
[0.5ex] \hline
$2\times10^{44}$  ~~~~~~~~~~~& $1.9$~~~~~~~~~~~ & $7.9\times10^{-11}$ \\
$3\times10^{44}$  ~~~~~~~~~~~& $1.85$~~~~~~~~~~~ & $7.9\times10^{-11}$ \\
$4\times10^{44}$  ~~~~~~~~~~~& $1.83$~~~~~~~~~~~ & $7.9\times10^{-11}$ \\[1ex]
\hline
\end{tabular}
\label{table:nonlin}
\end{table}

\subsection{Model III}

Using Eqs. (\ref{tb13}) and (\ref{z2}), we obtain the expression of
baryon to entropy ratio
\begin{eqnarray}\label{z3}
\frac{\eta_B}{S}&\simeq& \frac{\kappa^3\pi
\mathcal{T}_D^5g_b\sqrt{g_\ast}}{6\sqrt{10}M_\ast^2\gamma^{\frac{1}{2}}(\gamma-3f_1\gamma-f_1)^{\frac{3}{2}}}
\bigg((3\gamma-1)f\bigg(1+\ln\bigg(\frac{\gamma(3\gamma-1)\kappa^2\pi^2
\mathcal{T}_D^4g_\ast}{90\gamma(\gamma-3f\gamma-f)}\bigg)\bigg)-\gamma\bigg).
\end{eqnarray}
\begin{figure}
\centering \epsfig{file=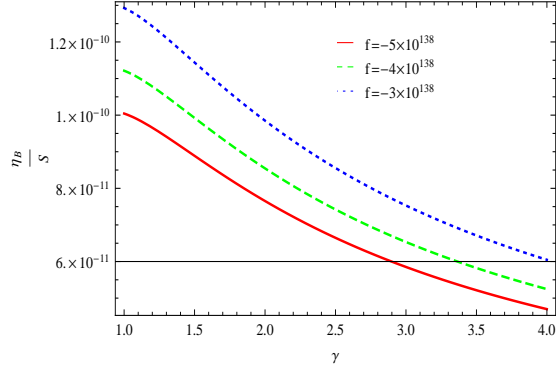, width=.45\linewidth,
height=2.1in} \caption{Plot of baryon to entropy ratio
$\frac{\eta_B}{S}$ as the function of parameter $\gamma$ in the
framework of generalized baryogenesis interaction for Model III, we
set the values of parameters as $g_b=1,~
\mathcal{T}_D=2\times10^{16},~M_\ast=10^{12},~g_\ast=106,~
\kappa=1$.}
\end{figure}
In Figure \textbf{7}, we plot $\gamma$-dependence of the baryon to
entropy ratio for different values of values of $f$. It informs us
that, for all values of $f$ by setting $\gamma=1.4,1.7,2.2$, we
obtain baryon to entropy ratio as
$\frac{\eta_B}{S}=9.2\times10^{-11}$, which satisfy the
observational constraints.
\begin{table}[]
\caption{Generalized Baryogenesis Interaction for $f(T,B)=-T+g(B)$}
\centering
\begin{tabular}{c c c}
\hline\hline
$f$ ~~~~~~~~~~~& $\gamma$~~~~~~~~~~~ & $\frac{\eta_B}{S}$ (Baryon to entropy ratio) \\
[0.5ex] \hline
$-5\times10^{138}$  ~~~~~~~~~~~& $1.4$~~~~~~~~~~~ & $9.2\times10^{-11}$ \\
$-4\times10^{138}$  ~~~~~~~~~~~& $1.7$~~~~~~~~~~~ & $9.2\times10^{-11}$ \\
$-3\times10^{138}$  ~~~~~~~~~~~& $2.2$~~~~~~~~~~~ & $9.2\times10^{-11}$ \\[1ex]
\hline
\end{tabular}
\label{table:nonlin}
\end{table}

\subsection{Model IV}

In the context of more complete generalized baryogenesis interaction
for this particular model,m we obtain the expression of baryon to
entropy ratio as
\begin{eqnarray}\nonumber
\frac{\eta_B}{S}&=&\frac{45\gamma g_b}{2\pi^2g_\ast M_\ast^2
\mathcal{T}_D}\left(\frac{\sqrt{9A_1^2\gamma^4+2\chi\left(\frac{\kappa^2\pi^2g_\ast
\mathcal{T}_D^4}{15}-A_0\right)}}{2\chi}+\frac{3A_1\gamma^2}{2\chi}\right)^{\frac{5}{2}}\bigg(12\gamma^3A_2\\\nonumber&+&
\frac{2A_1\chi\gamma}{3A_1\gamma^2+\sqrt{9A_1^2\gamma^4+2\chi\left(\frac{\kappa^2\pi^2g_\ast
\mathcal{T}_D^4}{15}-A_0\right)}}+18A_4\gamma^2(3\gamma-1)+24A_3\\\label{y1}&\times&\gamma(3\gamma-1)^2\bigg).
\end{eqnarray}
\begin{figure}
\centering \epsfig{file=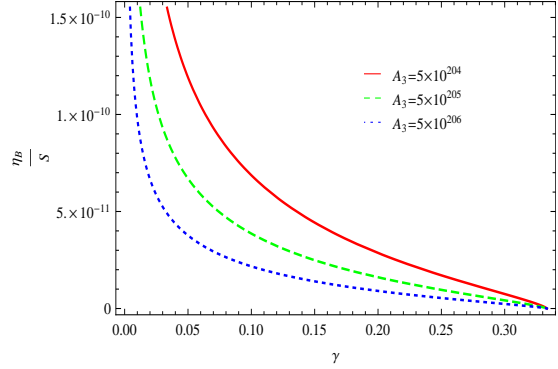, width=.45\linewidth,
height=2.1in} \caption{Plot of baryon to entropy ratio
$\frac{\eta_B}{S}$ versus parameter $\gamma$ in the light of
generalized baryogenesis interaction for Model IV. Other parameters
are $g_b=1,~\mathcal{T}_D=2\times10^{16},~
M_\ast=10^{12},~g_\ast=106,~\kappa=1,~A_0=2\times10^{10},~
A_1=3\times10^{10},~A_2=5\times10^{70}$ and $A_4=6\times10^{10}$.}
\end{figure}

It can be observed from Figure \textbf{8} that the baryon to entropy
ratio remains $\frac{\eta_B}{S}\leq 9\times10^{-11}$ for the range
of $\gamma\geq0.01$ which favors the observational bounds
\cite{1,2}. Detailed discussion is mentioned in the following
\textbf{Table VIII}.
\begin{table}[ht]
\caption{Generalized Baryogenesis interaction for $f(T,B)
=A_0+A_1T+A_2T^2+A_3B^2+A_4TB$} \centering
\begin{tabular}{c c c}
\hline\hline
$A_3$ ~~~~~~~~~~~& $\gamma$~~~~~~~~~~~ & $\frac{\eta_B}{S}$ (Baryon to entropy ratio) \\
[0.5ex] \hline
$5\times10^{204}$  ~~~~~~~~~~~& $0.07$~~~~~~~~~~~ & $8.99\times10^{-11}$ \\
$5\times10^{205}$  ~~~~~~~~~~~& $0.03$~~~~~~~~~~~ & $8.8\times10^{-11}$ \\
$5\times10^{206}$  ~~~~~~~~~~~& $0.01$~~~~~~~~~~~ & $8.8\times10^{-11}$ \\[1ex]
\hline
\end{tabular}
\label{table:nonlin}
\end{table}

\section{Conclusion}

This paper presented the detailed discussion of
gravitational baryogenesis mechanism in the context of $f(T,T_G)$
and $f(T,B)$ theories of gravity. For $f(T,T_G)$-gravity, we have
used two specific models $f(T,T_G)= \alpha_1 \sqrt{T^2+\alpha_2 T_G}
- T$ and $f(T,T_G)=\alpha_1 T^2+\alpha_2 T \sqrt{|T_G|}+ \beta_1
\sqrt{T^2+\beta_2 T_G}-T$. Similarly, we considered $f(T,B)=-T+g(B)$
where ($g(B)=f_1B \ln B$) and $f(T,B) =A_0+A_1T+A_2T^2+A_3B^2+A_4TB$
models in the framework of $f(T,B)$-gravity. For both theories of
gravity, we have chosen scale factor $a(t)=m_0t^\gamma$ and
constructed baryon to entropy ratio $\frac{\eta_B}{S}$ by assuming
that the universe filled by perfect fluid and dark energy. We also
evaluated more complete and generalized baryogenesis interaction
proportional to $\partial_\mu f(T+T_G)$ and $\partial_\mu f(T+B)$.
For all cases, our results have showed excellent consistency with
approximate observational value $\frac{\eta_B}{S}\sim 9.42 \times
10^{-11}$
\cite{1,2}. The core results of this work are given below. 

\begin{itemize}
  \item \textbf{Model I}: In Figure \textbf{1}, we show the plot of
  baryon to entropy ratio against parameter $\gamma$, which shows that observation value of baryon to entropy ratio
  can be met for $\gamma\leq2$ with $\alpha_2=10^{29}$.

  \item \textbf{Model II}: In Figure \textbf{2}, One can find the
  value of baryon to entropy ratio approximately equal to $7.5^{+1.5}_{-1.1}\times
  10^{-11}$ with $1.65\leq\gamma\leq 1.94$ for all cases of
  $\beta_1$, which satisfied the observational bounds.

  \item \textbf{Model III}: It is observed that $\gamma\leq1.56$,
  for all values of $f$, our result
  $\frac{\eta_B}{S}=7.5^{+1.5}_{-1.5}\times10^{-11}$ correspond to
  observationally measured value of baryon to entropy ratio (Figure \textbf{3}).

  \item \textbf{Model IV}: For this model, we observed (Figure
  \textbf{4})
  $5.5\times10^{-11}\leq\frac{\eta_B}{S}\leq8.09\times10^{-11}$, before
  $\gamma=2.5$ and
  $A_2=-2\times10^{22}$, which indicate the
  excellent agreement with observational value $\frac{\eta_B}{S}\sim 9.42 \times
  10^{-11}$. Anyhow, for other values of $A_2$ and $\gamma\geq1.25$,
  trajectories are very close to observational constraints.

In the following, we have given the results for the generalized
Baryogenesis interaction scenario. These are as follows:

  \item \textbf{Model I}: For this baryogenesis interaction (Figure \textbf{5}), for $\alpha_2=10^{81}$ and $1.15\lesssim \gamma\lesssim 1.5$,
  the ratio of baryon number density to entropy obtained by gravitational
  baryogenesis (\ref{ali3}) lies in the range  $2\times10^{-11}\lesssim  \frac{\eta_B}{S}\lesssim
  7.5\times10^{-11}$. While for $\alpha_2=2\times10^{81}$ and $1.15\lesssim \gamma\lesssim
  2$. this ratio correspond to $2\times10^{-11}\lesssim  \frac{\eta_B}{S}\lesssim
 9.4\times10^{-11}$. Similarly, for $\alpha_2=4\times10^{81}$ and $1.15\lesssim \gamma\lesssim
  2.5$, we have $2\times10^{-11}\lesssim  \frac{\eta_B}{S}\lesssim
 8.6\times10^{-11}$.

  \item \textbf{Model II}: For this model, the baryon to entropy
ratio at leading order is, $\frac{\eta_B}{S}=7.9\times10^{-11}$ for
all cases when $1.83\leq\gamma\leq1.9$, which is in very good
agreement with observations (Figure \textbf{6}).

  \item \textbf{Model III}: From the curves of the Figure
  \textbf{7}, we notice that for $\gamma=1.4,1.7,2.2$, implies
  $\frac{\eta_B}{S}=9.2\times10^{-11}$, which compatible with the observation
  data of baryon to entropy ratio.

  \item \textbf{Model IV}: For this model, our result is shown in
  Figure \textbf{8}, which provides a well matched observational
  value when $\gamma\geq0.01$.
\end{itemize}

\end{document}